\documentclass[review]{elsarticle}

\usepackage{lineno,hyperref}
\modulolinenumbers[5]

\journal{ Journal of Communications and Networks }

\usepackage[utf8]{inputenc}
\usepackage{tikz}
\usepackage{amsfonts}
\usepackage{amsmath}
\usepackage{mathtools}
\usepackage{colortbl}
\usepackage{diagbox}
\usepackage{multirow}
\usepackage{fixltx2e}
\usepackage{graphicx}

\usepackage{setspace}
\usepackage[T2A]{fontenc}
\usepackage[russian,english]{babel}
\usepackage[babel=true]{microtype}
\usepackage{algorithm}
\usepackage{algpseudocode}
\usepackage{physics}
\usepackage{booktabs} % For professional looking tables
\usepackage{geometry}

\usepackage{adjustbox}

\begin{document}
	
	\begin{frontmatter}
		
		\title{On the effect of the average clustering coefficient on topology-based link prediction in featureless graphs}

		\author[mymainaddress]{Mehrdad Rafiepour}
		\ead{mehrdad.rafie.p@gmail.com}
		\author[mymainaddress]{S. Mehdi Vahidipour\corref{mycorrespondingauthor}}
		\ead{vahidipour@kashanu.ac.ir}
		\address[mymainaddress]{Computer and Electrical Engineering Department, University of Kashan, Kashan, Iran.}
		\cortext[mycorrespondingauthor]{Corresponding author}
		\begin{abstract}
			Link prediction is a fundamental problem in graph theory with diverse applications, including recommender systems, community detection, and identifying spurious connections. While feature-based methods achieve high accuracy, their reliance on node attributes limits their applicability in featureless graphs. For such graphs, structure-based approaches, including common neighbor-based and degree-dependent methods, are commonly employed. However, the effectiveness of these methods depends on graph density, with common neighbor-based algorithms performing well in dense graphs and degree-dependent methods being more suitable for sparse or tree-like graphs. Despite this, the literature lacks a clear criterion to distinguish between dense and sparse graphs. This paper introduces the average clustering coefficient as a criterion for assessing graph density to assist with the choice of link prediction algorithms. To address the scarcity of datasets for empirical analysis, we propose a novel graph generation method based on the Barabasi-Albert model, which enables controlled variation of graph density while preserving structural heterogeneity. Through comprehensive experiments on synthetic and real-world datasets, we establish an empirical boundary for the average clustering coefficient that facilitates the selection of effective link prediction techniques. 
		\end{abstract}
		
		
	\end{frontmatter}
	
	%\linenumbers
	
	\section{Introduction}
		Link prediction has always been a challenging problem in graph theory due to its many applications in various problems, including recommender systems \citep{RecommenderSystems}, detection of unformed connections  \citep{CommunityLP}, or finding spurious links \citep{NoisyLinks}.
		
		Methods for link prediction often rely on node features to detect connections \citep{AttributeLP}. While these approaches achieve high accuracy, their reliance on node features makes them unsuitable for featureless graphs. Several challenges arise from this dependence. First, obtaining feature-rich graphs can be expensive, both in terms of financial cost and time investment. Second, performing link prediction on feature-rich graphs typically demands high-performance hardware, leading to increased computational costs. Third, certain graphs contain sensitive data, such as users' private information, and exposing such information to link prediction systems raises significant privacy concerns \citep{PrivacyConcern}. Fourth, some graphs inherently contain no features at all. In these scenarios, link prediction focuses on analyzing the graph's structural properties, such as node degrees, edge counts, and edge connectivity arrangement  \citep{TopologyAwareLP}.
		
		Common neighbor-based and degree-dependent algorithms are two of the most prominent methods of link prediction. Jaccard \citep{Jaccard} and Adamic-Adar \citep{AdamicAdar} are two examples of common neighbor algorithms. These algorithms focus on analyzing the existence of the closed triplets in graphs. For example, suppose that in a network, nodes represent individuals, and the link between two nodes represents a friendship between two individuals. In this case, if two individuals have one or more common friends, the probability of forming a friendship relationship between them in the future increases significantly. However, the effective use of this method requires the presence of a sufficient number of links in the network structure.
		In contrast, the Heterogeneity and Homogeneity Index are two instances of degree-dependent methods that utilize the degree of nodes \citep{HEIHOI}.
		
		Common neighbor-based algorithms perform well in dense graphs, where the number of edges is close to the maximum possible number. However, the number of closed triplets dwindles with a decreased number of edges, making common neighbor-based methods falter since they rely on closed triplets for accurate link prediction. Hence, in tree-like networks, where the number of edges is close to $N+1$ for $N$ nodes, they cannot be utilized. Similarly, sparse graphs with a minimal number of closed triplets cannot be processed with common neighbor-based algorithms. To this end, the Homogeneity Index and the Heterogeneity Index are used to predict links in empty or tree-like graphs \citep{HEIHOI}.
		
		In the absence of features for nodes and edges, graph structure-based methods are most useful. However, it is unclear when node degree-based methods can be used instead of common neighbor-based methods. In other words, the literature does not define a criterion to discern dense and tree-like graphs, a problem vital to selecting an appropriate topology-based link prediction algorithm. Therefore, in this paper, we investigate the role of the average clustering coefficient as a suitable criterion for determining the density or emptiness of a graph by conducting empirical experiments.
		
		Since the number of available datasets is insufficient to create an empirical boundary, in this paper, we propose a new graph alteration method based on the Barabasi-Albert approach \citep{BarabasiAlbert}. To confirm the hypothesis of this paper, we conduct several experiments by calculating the average clustering coefficient, Homogeneity-Heterogeneity Index, and Jaccard score on known datasets and present the results according to the AUC value \citep{AUC}.
		In summary, the innovations of this paper are fourfold:
		
		\begin{enumerate}
			\item We investigate the feasibility of using the average clustering coefficient as a criterion to identify the best link prediction algorithm.
			\item We define a novel graph generation approach based on the Barabasi-Albert graph to increase the average clustering coefficient while preserving graph heterogeneity.
			\item Leveraging the proposed algorithm, we define an average clustering coefficient boundary, according to which we can find the appropriate link detection score.
			\item We assess the validity of the proposed criterion and the proposed graph generation method used to calculate the boundary with real-world datasets. 
		\end{enumerate}
		The remainder of this paper is organized as follows: Section \ref{definitions} defines the key terms and jargon used in this study. Section \ref{problem} outlines the problem under investigation. The proposed methodology is described in section \ref{method}. To assess the effectiveness of the proposed methodology, empirical experiments are conducted in section \ref{experiments}, followed by a discussion of the results. Finally, the article concludes in Section \ref{conclusion}.
	\section{Definitions}\label{definitions}
		In order to delineate the problem, this section sets forth definitions of the primary components of the discussed issue.
		\subsection{Link Prediction}
			Graph $G=(V,E)$ is defined by a set of nodes $V$ and a set of edges $E$ that are unordered pair of nodes $e = (v_i, v_j) \in E$, where $v_i, v_j \in V$. Each pair $(v_i, v_j)$ can appear in at most one edge $e$.
			In link prediction, the observed set of edges $ E $ is divided into $E^T$ and $E^P$, such that $E^T \cup E^P = E$ and $E^T \cap E^P = \emptyset$. Links in $E^T$ are used to train or determine the algorithm, while the links in $ E^P $ are used to evaluate the link prediction algorithm. Moreover, the set of unobserved edges, which is the set of edges that are not in $E$ but are possible combinations of nodes inside $V$, is defined by $\bar{E}$.	
		\subsection{Link Prediction With Heterogeneity and Homogeneity Index}
			The Heterogeneity and Homogeneity Indexes assume there is a correlation between the degree of a node and its potential to connect to its neighbors, a behavior observed in social networks \citep{HEIHOI}. Accordingly, the Heterogeneity Index assigns the highest score to the node with the greatest difference. In contrast, the maximum score in the Homogeneity Index is given to the node with the lowest degree difference between the two nodes. 
			\subsubsection{Heterogeneity Index}
				This algorithm is defined by $S_{ij}^{HEI}$ in Equation \ref{HEIFormula}, where $\alpha$ is the heterogeneity exponent, and $k(i), k(j) $ represent the degrees of nodes $i$ and $j$, respectively. The $S^{HEI}$ rewards higher degree heterogeneity.
				\begin{equation}\label{HEIFormula} 
					S_{ij}^{HEI}= \abs{k(i)-k(j)}^{ \alpha } ,   {\scriptstyle0 \leq \alpha \leq 1} 
				\end{equation}
			\subsubsection{Homogeneity Index}
				$S_{ij}^{HOI}$ in Equation \ref{HOIFormula} increases degree homogeneity, allocating higher prediction scores to nodes with similar orders.
				\begin{equation}\label{HOIFormula} 
					S_{ij}^{HOI}= \dfrac{1}{\abs{k(i)-k(j)}^{ \alpha }}  ,   {\scriptstyle0 \leq \alpha \leq 1} 
				\end{equation}
		\subsection{Jaccard Similarity Score}
			The Jaccard score is a traditional and widely recognized metric used in link prediction, which is defined based on the concept of common neighbors in graphs. This score specifically measures the ratio of the number of common neighbors between two nodes to the total number of unique neighbors of those nodes. Equation \ref{JaccardFormula} represents the Jaccard score between nodes $i$ and $j$ with the notation $S_{ij}^{JAC}$, where Ґ$(i)$ and Ґ$(j)$ denote the set of neighbors of node $i$ and $j$, respectively.
			\begin{equation}\label{JaccardFormula} 
				S_{ij}^{JAC}= \dfrac{\abs{ \text{Ґ} (i) \cap (j)}}{\abs{ \text{Ґ}(i) \cup \text{Ґ}(j)}} 
			\end{equation}
		\subsection{Average Clustering Coefficient}
			The average clustering coefficient\citep{clusteringcoefficient} calculates the average interconnectivity of nodes inside a graph. The clustering coefficient is defined as the ratio of closed triplets to total triplets, both open and closed. $C$ in Equation \ref{averageclusteringcoefficientFormula} below represents the average clustering coefficient.
			\begin{equation}\label{averageclusteringcoefficientFormula}
				c_i = \frac{2 \times \text{Number of triangles of node } i}{k_i (k_i - 1)}, \quad C = \frac{1}{N} \sum_{i=1}^{N} c_i
			\end{equation}
		\subsection{Barabasi-Albert Graph}
		The Barabasi-Albert model \citep{BarabasiAlbert} is a scale-free graph generation algorithm with a preferential dependency mechanism. In this graph, nodes with unusually high degrees are generated. The Barabasi-Albert method starts with an initially connected graph containing $m_{initial}$ nodes. In each step, a single node joins the graph, and $m, m \geq m_{initial}$ links are added based on the probability described in Equation \ref{BarabasiAlbertFormula}, where $K_{i}$ is the degree of node $i$ and $p_{i}$ represents the probability that a link is formed with node $i$. 
		\begin{equation}\label{BarabasiAlbertFormula} 
			p_{i}=\dfrac{k_{i}}{ \Sigma_{j} k_{j}}
		\end{equation}
		\subsection{Area Under the Receiver Operating Characteristic Curve}
			We use Area Under the Receiver Operating Characteristic Curve(AUC), as defined in Equation \ref{aucformula}, for evaluation. To assess the performance, $m$ random edges $e_{1}\in E^{P}$ and $e_{2}\in \bar{E}$ are selected, and their respective scores are calculated based on the chosen algorithm $Score$. Accordingly, $m'$ represents the number of times where $Score_{e_{1}} > Score_{e_{2}}$ and $m''$ is the  number of times  $Score_{e_{1}} = Score_{e_{2}}$. 
			\begin{equation}\label{aucformula} 
				AUC = \dfrac{m'+\dfrac{1}{2} m''}{m}
			\end{equation}
	\section{Problem Definition}\label{problem}
			The main question in link prediction is choosing the appropriate algorithm according to the graph structure. So far, no specific boundary has been introduced to determine whether the network is dense or sparse enough, raising challenges in algorithm selection. To resolve this ambiguity, we employ the average clustering coefficient and conduct empirical tests on a proposed artificial graph to establish a boundary that delineates the conditions under which traditional common neighbor-based criteria are most effective and when homogeneity and heterogeneity scores are more appropriate. Moreover, we test the contrived boundary on numerous real-world graphs to assess the boundary's validity.
	\section{Proposed Methodology}\label{method}
		We hypothesize that with an increased clustering coefficient, common neighbor-based algorithms work more effectively. Since there are not enough graphs to reach an average clustering coefficient boundary, we introduce an artificial graph alteration method in the present section. Initially, we create a tree graph using the Barabasi-Albert method with $m=1$. This tree contains zero closed triangles as each new node exactly connects to one existing node with a probability distribution skewed towards higher-degree nodes, creating a heterogeneous graph. Second, we add edges to the initialized graph in a way that allows us to increase the clustering coefficient while preserving the heterogeneity of the graph.
		\subsection{Proposed Graph Construction Method}\label{proposedmethodsection} 
			A random node $i \in V$ in a Barabasi-Albert graph $G=(V, E)$ with $m=1$ is selected (order does not matter). In the next step, a random node $z \in \text{Ґ}_{i}$ in $i$'s list of neighbors $\text{Ґ}_{i} \subset V$ is chosen. In $z$'s neighbor list $ \text{Ґ}_{z}$, a node $j \in \text{Ґ}_{z}$ is picked with respect to the probability distribution formulated in Equation \ref{MethodFormula}. Lastly, a new edge $e_{(i,j)}$ with the probability $d$ is added to the graph's edge list.
			\\
			In summary, initial node $i$ can form a new edge $e_{(i,j)}$ with the probability $d$ with one of its neighbor's neighbors, forming a new closed triangle. This surge of closed triangles increases the value of the average clustering coefficient. 
			\\
			To construct a triangle, node $i$ must select one of the neighbors of node $z$ in such a way that it has the highest probability of connecting to the node that has the largest difference in degree from node $i$. We ensure this by introducing Equation \ref{MethodFormula}. Given that the proposed approach is stochastic, candidates connecting to $i$ are selected probabilistically to not affect other graph properties.
			\\
			In Equation \ref{MethodFormula}, $P_{select}(i,j)$ is the probability that $j$ is picked for $i$ and $K(i)$ specifies the degree of node $i$. 
			\begin{equation}\label{MethodFormula} 
				P_{Select}(i,j)= \dfrac{\abs{K(i)-K(j)}}{\Sigma_{z' \in \text{Ґ}_{(z)}} \abs{K(i)-K(z')}} , {\scriptstyle i \in V, j \in \text{Ґ}_{(z)}} 
			\end{equation}
			Algorithm \ref{AlgorithmProposed} describes the proposed method, where $\epsilon$ is a small number and $WeightedSelection$ is a function that selects a single item in a list $\text{Ґ}_{z}$ based on given probability distribution $Probability$. 
			\begin{algorithm}
				\caption{Proposed Graph Alteration Algorithm}\label{AlgorithmProposed} 
				\begin{algorithmic}[1]
					\Procedure{Populate}{d}
					\State {Initialize graph $G=(E, V) \gets BarababasiAlbert(m=1)$}
					\For{$i$ in $V$}
					\For{$z$ in $\text{Ґ}_{i}$}
					\State{$DifferenceSum \gets \Sigma_{z' \in \text{Ґ}_{(z)}} \abs{ K(i) - K(z')+ \epsilon}$}
					\State{Initialize dict $Probability$}
					\For{$j$ in $\text{Ґ}_{z}$}
					\State $Probability[j]$ =  $\dfrac{\abs{ K(i) - K(j) + \epsilon }} { DifferenceSum }$
					\EndFor
					\State{$EdgeSelection = WeightedSelection(\text{Ґ}_{z},Probability,d)$}
					\State{$E \gets EdgeSelection$}
					\EndFor
					\EndFor
					\State \textbf{return} $G$
					\EndProcedure
				\end{algorithmic}
			\end{algorithm}

	\section{Experiments}\label{experiments}
		In this section, four experiments are conducted to investigate the relationship between the value of the average clustering coefficient with network characteristics and the performance of degree-dependent and common neighbor-based link prediction algorithms based on AUC. In this set of experiments, the number of graph order is considered variable, and the $\alpha$ equals $0.5$.
		To compare and evaluate the efficiency of the algorithms, the Jaccard score is used as a representative of the common neighbor-based algorithms.
		\subsection{The Effect of $d$ on Average Clustering Coefficient}
\begin{figure*}
			\includegraphics{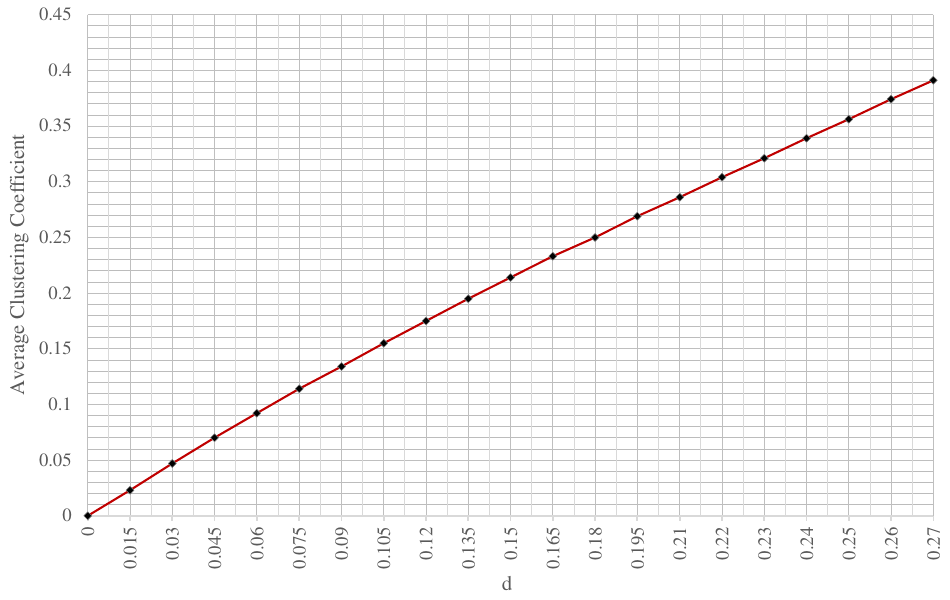}
			\caption{\label{dvsavcc} The effect of $d$ on Average Clustering Coefficient $AvCC$.}
		\end{figure*}

		To find the appropriate value of d for generating a specific clustering coefficient, we measured the effect of varying d on the generated clustering coefficient. The results of this experiment are shown in Figure \ref{dvsavcc}. In this experiment, for a graph with a fixed number of nodes $N=100,000$, it is observed that the clustering coefficient increases relatively linearly with increasing probability $d$ of forming a closed triangle. Also, for $d=0$, the clustering coefficient is zero, as the Barabasi-Albert graph is generated with $m=1$. With $d=0$, no closed triplets can form, so the clustering coefficient is zero per Equation \ref{averageclusteringcoefficientFormula}.
		\subsection{Clustering Coefficient Boundary}
\begin{figure*}
				\includegraphics{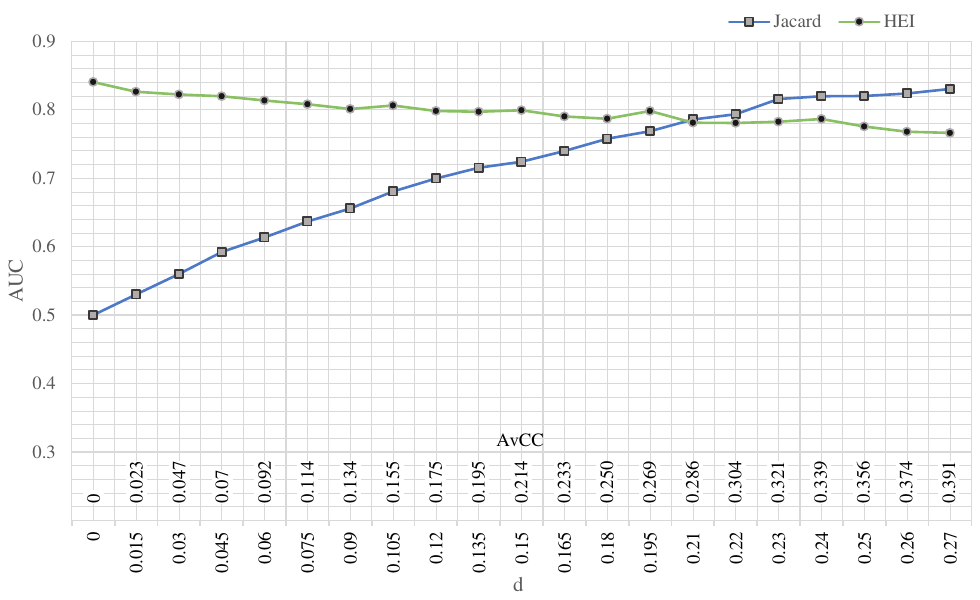}
				\caption{\label{scorecontest} AUC results for $S^{HEI}$ and $S^{JAC}$ with rise in the value of $d$ and subsequently the Average Clustering Coefficient denoted by $AvCC$.}
			\end{figure*}

			As evident in Figure \ref{dvsavcc}, with an increase in the probability of closed triplets forming, the clustering coefficient has an upward trend. In Figure \ref{scorecontest}, the behavior of degree-dependent and the common neighbor-based algorithms with changes in $d$ is compared. As shown in Figure \ref{scorecontest}, with a rise in the value of $d$, the performance of $S^{HEI}$ almost stays on the same level while the performance of $S^{JAC}$ improves significantly as the graph gradually becomes dense. At $d = 0.21$, $S^{JAC}$ performance surpasses $S^{HEI}$ and obtains a better $AUC$. Based on these observations, it can be inferred that in sparse graphs whose clustering coefficient is less than approximately $0.27$, the degree-dependent algorithms have a higher accuracy in link prediction. In graphs with a clustering coefficient between $0.2$ and $0.3$, the difference in accuracy between the algorithms is close. As a result, as the graph becomes denser and the number of closed triplets increases, degree-dependent algorithms are no longer superior to common neighbor-based ones.
		\subsection{Boundary's dependence on Graph's Order}
\begin{figure*}
				\includegraphics{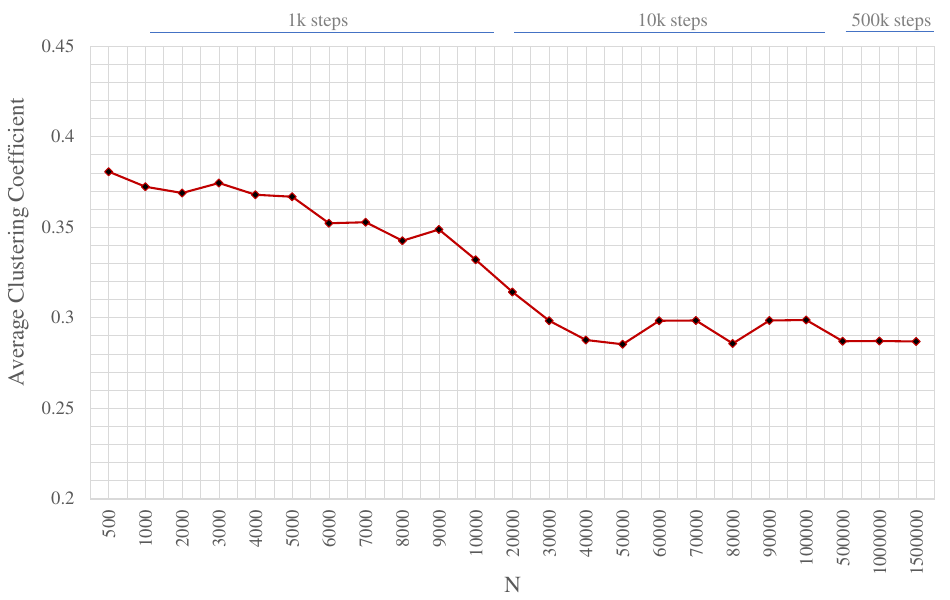}
				\caption{\label{graphorder} Changes in the average clustering coefficient with changes in the number of nodes $N$ in a graph.}
			\end{figure*}

			In this experiment, we investigated the effect of the graph's order on the clustering coefficient boundary. As shown in Figure \ref{graphorder}, the size of the clustering coefficient depends on the number of nodes $N$, and as $N$ goes up, the clustering coefficient gradually decreases and stabilizes around the approximate value of $0.3$ for $100,000$ nodes. The results of this experiment show that the selection of an appropriate criterion for link prediction depends on the number of graph nodes. Note that $d$ varied between $0.23$ and $0.3$ in our experiments.
		\subsection{Validating the Boundary with Real Graphs}
			In order to assess the accuracy of the obtained clustering coefficient boundary, we calculated the average clustering coefficient and compared the performance of the Jaccard score with degree-dependent algorithms using $AUC$ for a number of real-world graphs. The results of this comparison are described in Table \ref{realtable}. Experiments confirm the approximations of the proposed algorithm and show that for clustering coefficiency less than 0.3, $S^{HEI}$ and $S^{HOI}$ perform better than $S^{JAC}$. However, as the clustering coefficient approaches 0.3, the performance of $S^{JAC}$ gradually improves, and the difference in $AUC$ with  $S^{HEI}$ and $S^{HOI}$ decreases. Similarly, as the clustering coefficient exceeds $0.3$, the performance of  $S^{HEI}$ and $S^{HOI}$ tends to decline and perform worse due to the increase in closed triplets, while $S^{JAC}$ reaches an $AUC$ close to 1.
			
			\begin{table}[ht!]
				\centering
				\caption{\label{realtable}The $AUC$ Performance comparison of $S^{HEI}$, $S^{HOI}$, and $S^{JAC}$ on real graphs, sorted by $AvCC$ (low to high). $AvCC$ denotes the Average Clustering Coefficient, while $N$ and $E$ represent the number of nodes and number of edges, respectively.}
				\begin{tabular}{lccccccc}
					\toprule
					\textbf{$\mathbf{Graph Name}$} & \textbf{$\mathbf{N}$}   & \textbf{$\mathbf{E}$} & \textbf{$\mathbf{AvCC}$} & \textbf{$\mathbf{S^{HOI}}$} & \textbf{$\mathbf{S^{HEI}}$} & \textbf{$\mathbf{S^{JAC}}$} \\
					\midrule
					Gnutella deer to deer network & 62586 & 147892 & 0.0054 & 0.3347 & $\mathbf{0.6653}$ & 0.5184 \\
					Mouse Brain (ogbl-vessel)\citep{vessel}  & 3463841 & 4276719 & 0.0151 & $\mathbf{0.5460}$ & 0.4541 & 0.4855 \\
					SNAD/Higgs-twitter & 256491 & 327374 & 0.0156 & 0.0584 & $\mathbf{0.9416}$ & 0.5325 \\
					Road network of California & 1965206 & 2766607 & 0.0463 & $\mathbf{0.5707}$ & 0.4293 & 0.5620 \\
					Road network of Pennsylvania & 1088092 & 1541898 & 0.0464 & $\mathbf{0.5722}$ & 0.4278 & 0.5645 \\
					Road network of Texas & 1379917 & 1921660 & 0.0470 & $\mathbf{0.5661}$ & 0.4339 & 0.5560 \\
					Slashdot social network & 82168 & 504230 & 0.0603 & 0.0967 & $\mathbf{0.9033}$ & 0.7375 \\
					Who-trusts-whom network of Edinions.com & 75879 & 405740 & 0.1377 & 0.1009 & $\mathbf{0.8991}$ & 0.8798 \\
					GitHub-social & 37700 & 289003 & 0.1675 & 0.1432 & $\mathbf{0.8568}$ & 0.8038 \\
					Protein Associations (ogbn-proteins)\citep{proteingraph} & 132534 & 39561252 & 0.2798 & 0.4616 & 0.5432 & $\mathbf{0.9908}$ \\
					Amazon product co-purchasing & 403394 & 2443408 & 0.4176 & 0.3637 & 0.6363 & $\mathbf{0.9455}$ \\
					Drug-drug Interaction (ogbl-ddi)\cite{drugdruggraph} & 4267 & 1067911 & 0.5142 & 0.5347 & 0.4658 & $\mathbf{0.9469}$ \\
					Ego-facebook & 4039 & 88234 & 0.6055 & 0.4946 & 0.5054 & $\mathbf{0.9915}$ \\
					Author Collaborations (ogbl-collab)\citep{authorgraph} & 235868 & 967632 & 0.7290 & 0.3305 & 0.6651 & $\mathbf{0.9787}$ \\
					\bottomrule
				\end{tabular}
			\end{table}
		\subsection{Validation on Open Graph Benchmark}
			To further assess the validity of our findings, we evaluated the performance of the clustering coefficient using the Open Graph Benchmark \citep{ogb}. This benchmark includes tasks spanning biology, academic networks, and knowledge graphs, enabling researchers to test the performance of graph learning models and algorithms on real-world data.\\
			We specifically use the open graph benchmark with two goals in mind. The first goal is to add more datasets to an already extensive roster we employed to validate the boundary found, and the second is to evaluate the role of the boundary on other performance metrics such as 100-hit, 50hit and 20hit. Table \ref{obgtable} compares the link prediction performance of $S^{HEI}$, $S^{HOI}$, and $S^{JAC}$ on three undirected graphs. The $AvCC$ boundary is also valid using the hit metric, where for high values of AvCC, the hit value of datasets for $S^{JAC}$ is higher than $S^{HOI}$ and $S^{HEI}$. In contrast, the value of latter algorithms are higher in low AvCC values compared to the former. Notably, $S^{HOI}$ demonstrated superior performance compared to most proposed models on $Vessel$, a graph derived from the mouse brain. This result is particularly striking given that $S^{HOI}$ operates with zero learning parameters and exhibits very low computational complexity. We compared the performance of $S^{HOI}$ to existing learning-based methods in Section \ref{appendixvessel}.
			\begin{table}[h!]
				\centering
				\caption{\label{obgtable} Boundary evaluation with different metrics on open graph benchmark. $N$ represents the order of the graph, while $AvCC$ is the average clustering coefficient.}
				\begin{adjustbox}{max width=\textwidth}
					\begin{tabular}{@{}lccclcc@{}}
						\toprule
						\textbf{$\mathbf{Graph Name}$} & \textbf{$\mathbf{N}$} & \textbf{$\mathbf{AvCC}$} & \textbf{$\mathbf{Metric}$} & \textbf{$\mathbf{Type}$} & \textbf{$\mathbf{Validation (\%)}$} & \textbf{$\mathbf{Test (\%)}$} \\
						\midrule
						\multirow{3}{*}{\parbox[c]{5cm}{Mouse Brain\\(ogbl-vessel)\citep{vessel}}} & \multirow{3}{*}{\centering 3463841} & \multirow{3}{*}{\centering 0.0151} & \multirow{3}{*}{ROC-AUC} & $S^{JAC}$ & 48.55 & 48.54 \\
						& & & & $S^{HEI}$ & 45.41 & 45.40 \\
						& & & & $S^{HOI}$ & 54.59 & 54.60 \\
						\midrule
						\multirow{9}{*}{\parbox[c]{5cm}{Drug-Drug Interaction\\(ogbl-ddi)\citep{drugdruggraph}}} & \multirow{9}{*}{\centering 4267} & \multirow{9}{*}{\centering 0.5142} & \multirow{3}{*}{Hits@20} & $S^{JAC}$ & 0.00 & 0.00 \\
						& & & & $S^{HEI}$ & 0.00 & 0.00 \\
						& & & & $S^{HOI}$ & 0.00 & 0.00 \\
						\cmidrule(lr){4-7}
						& & & \multirow{3}{*}{Hits@50} & $S^{JAC}$ & 0.01 & 0.07 \\
						& & & & $S^{HEI}$ & 0.01 & 0.00 \\
						& & & & $S^{HOI}$ & 0.00 & 0.00 \\
						\cmidrule(lr){4-7}
						& & & \multirow{3}{*}{Hits@100} & $S^{JAC}$ & 5.54 & 4.86 \\
						& & & & $S^{HEI}$ & 0.02 & 0.00 \\
						& & & & $S^{HOI}$ & 0.00 & 0.00 \\
						\midrule
						\multirow{9}{*}{\parbox[c]{5cm}{Authors Collaboration\\(ogbl-collab)\citep{authorgraph}}} & \multirow{9}{*}{\centering 235868} & \multirow{9}{*}{\centering 0.7290} & \multirow{3}{*}{Hits@20} & $S^{JAC}$ & 38.20 & 29.13 \\
						& & & & $S^{HEI}$ & 1.34 & 1.34 \\
						& & & & $S^{HOI}$ & 0.00 & 0.00 \\
						\cmidrule(lr){4-7}
						& & & \multirow{3}{*}{Hits@50} & $S^{JAC}$ & 60.98 & 50.50 \\
						& & & & $S^{HEI}$ & 2.63 & 2.74 \\
						& & & & $S^{HOI}$ & 0.00 & 0.00 \\
						\cmidrule(lr){4-7}
						& & & \multirow{3}{*}{Hits@100} & $S^{JAC}$ & 64.89 & 54.54 \\
						& & & & $S^{HEI}$ & 4.44 & 4.33 \\
						& & & & $S^{HOI}$ & 0.00 & 0.00 \\
						\midrule
						\multirow{9}{*}{\parbox[c]{5cm}{Protein-Protein Association\\(ogbl-ppa)\citep{proteingraph}}} & \multirow{9}{*}{\centering 576289} & \multirow{9}{*}{\centering 0.2228} & \multirow{3}{*}{Hits@20} & $S^{JAC}$ & 1.78 & 1.75 \\
						& & & & $S^{HEI}$ & 0.06 & 0.05 \\
						& & & & $S^{HOI}$ & 0.00 & 0.00 \\
						\cmidrule(lr){4-7}
						& & & \multirow{3}{*}{Hits@50} & $S^{JAC}$ & 12.20 & 9.07 \\
						& & & & $S^{HEI}$ & 0.07 & 0.07 \\
						& & & & $S^{HOI}$ & 0.00 & 0.00 \\
						\cmidrule(lr){4-7}
						& & & \multirow{3}{*}{Hits@100} & $S^{JAC}$ & 20.68 & 20.85 \\
						& & & & $S^{HEI}$ & 0.09 & 0.08 \\
						& & & & $S^{HOI}$ & 0.00 & 0.00 \\
						\bottomrule
					\end{tabular}
				\end{adjustbox}
			\end{table}
		\subsection{Leaderboard for OGBL-Vessel\label{appendixvessel}}
			Table \ref{vessel_leaderboard} presents a summarized leaderboard of the ogbl-vessel graph\footnote{The data in this table was retrieved on January 9, 2025, from \url{https://ogb.stanford.edu/docs/leader_linkprop/\#ogbl-vessel}}. Here, $\#Param$ represents the number of learnable parameters, $Hardware$ specifies the type and model of the hardware used for computation, and Valid and Test scores are evaluated using the $ROC-AUC$ criterion. The Homogeneity Index, Heterogeneity Index, and Jaccard Index, which are highlighted in bold, are $S^{HOI}$, $S^{HEI}$, and $S^{JAC}$, respectively. The results indicate that the Homogeneity Index performs better than all heuristic methods and some learning-based methods. 
			\begin{table}[ht]
				\centering
				\caption{\label{vessel_leaderboard}Performance of different methods based on Test and Validation ROC-AUC for the Vessel graph(ogbl-vessel).}
				\begin{tabular}{@{}cccccc@{}}
					\toprule
					$\mathbf{Rank}$ & $\mathbf{Method}$ & $\mathbf{Test}$      & $\mathbf{Valid}$     & $\mathbf{\#Params}$    & $\mathbf{Hardware}$        \\ \midrule
					1               & GAV\citep{gav}               & 0.9838               & 0.9837               & 8,194                  & V100 (32 GB GPU)           \\
					2               & SIEG\citep{SIEG}              & 0.8249               & 0.8255               & 594,498                & P100 (16GB GPU)                \\
					3               & SEAL\citep{SEAL}              & 0.8050               & 0.8053               & 172,610                & RTX 8000 Ti (48GB GPU)     \\
					4               & $\mathbf{Homogeneity\ Index}$ & $\mathbf{0.5460}$ & $\mathbf{0.5459}$    & $\mathbf{0}$           & $\mathbf{CPU\ (32GB\ RAM)}$ \\
					5               & LRGA\citep{LRGA}              & 0.5415               & 0.5418               & 26,577                 & A100 (80GB GPU)                \\
					6               & Matrix Factorization\citep{ogb} & 0.4997             & 0.4999               & 8,641                  & RTX 8000 Ti (48GB GPU)     \\
					7               & $\mathbf{Jaccard}$ & $\mathbf{0.4854}$    & $\mathbf{0.4855}$    & $\mathbf{0}$           & $\mathbf{CPU\ (32GB\ RAM)}$ \\
					8              & GCN w/Node2Vec\citep{ogb}    & 0.4954               & 0.4960               & 226,744,513            & RTX 8000 Ti (48GB GPU)     \\
					9              & Common Neighbors\cite{ogb}  & 0.4849               & 0.4850               & 0                      & RTX 8000 Ti (48GB GPU)     \\
					10              & MLP\citep{ogb}               & 0.4794               & 0.4801               & 1,037,577              & RTX 8000 Ti (48GB GPU)     \\
					11              & $\mathbf{Heterogeneity\ Index}$ & $\mathbf{0.4540}$ & $\mathbf{0.4541}$    & $\mathbf{0}$           & $\mathbf{CPU\ (32GB\ RAM)}$ \\
					12              & GCN\citep{ogb}               & 0.4353               & 0.4349               & 396,289                & RTX 8000 Ti (48GB GPU)     \\ \bottomrule
				\end{tabular}
			\end{table}

	\section{Conclusion}\label{conclusion}
		In this paper, we investigated the role of the clustering coefficient on structure-based link prediction. Our experiments showed that a boundary of approximately 0.27 to 0.37 separates the two degree-dependent and common neighbor-based algorithms in such a way that for a clustering coefficient higher than this value, common neighbor-based methods can be used with high confidence, and for a clustering coefficient lower than that, degree-dependent criteria are a reasonable choice. To obtain this boundary, we proposed a method based on the Barabasi-Albert graph and tested the boundary obtained by the proposed method with real-world data. The experimental results indicated the validity of this paper's hypothesis. Our results indicate that the clustering coefficient can be a good indication for detecting an appropriate link prediction algorithm.
		
		\subsection{Limitations}
			Our study focused on the empirical analysis of graphs. Therefore, the criterion introduced is not to be considered concrete. There are extreme cases where these observations might not hold. For example, a primarily sparse graph with one clump of extremely dense interconnected nodes will have a relatively low average clustering coefficient, while common neighbor-based methods will mostly work.
			Moreover, we assume graphs have degree-dependent property that allows link prediction, such as heterogeneity and homogeneity. However, this assumption might not always be true, and the degree of distribution may be random. Despite these limitations, we showed that the clustering coefficient is a reasonable criterion for knowing that at least a common neighbor-based method will work after a certain clustering threshold.
		
	\bibliography{custom}
	\appendix

\end{document}